\documentclass[3p,preprint,12pt]{elsarticle}
\usepackage{amsmath,amsfonts,amssymb}

\usepackage{graphicx}
\usepackage{color}

\def\bm{\boldsymbol}

\journal{Wave Motion}

\begin{document}
\begin{frontmatter}

\title{Extended shallow water wave equations}

\author[uoi]{Theodoros P. Horikis}
\address[uoi]{Department of Mathematics, University of Ioannina,
            Ioannina
            45110,
            Greece}

\author[uoa]{Dimitrios J. Frantzeskakis}
\address[uoa]{Department of Physics, University of Athens,
            Panepistimiopolis, Zografos,
            Athens
            15784,
            Greece}

\author[edin,wollo]{Noel F. Smyth}
\address[edin]{School of Mathematics, University of Edinburgh,
            Edinburgh
            EH9 3FD,
            Scotland,
            U.K.}
\address[wollo]{School of Mathematics and Applied Statistics,
University of Wollongong,
            Northfields Avenue,
            Wollongong
            2522,
            New South Wales,
            Australia}

\begin{abstract}
Extended shallow water wave equations are derived, using the method of asymptotic
expansions, from the Euler (or water wave) equations. These extended models are valid one order
beyond the usual weakly nonlinear, long wave approximation, incorporating all appropriate
dispersive and nonlinear terms. Specifically, first we derive the extended Korteweg-de Vries
(KdV) equation, and then proceed with the extended Benjamin-Bona-Mahony and the extended
Camassa-Holm equations in $(1+1)$-dimensions, the extended cylindrical KdV equation in the
quasi-one dimensional setting, as well as the extended Kadomtsev-Petviashvili and its cylindrical
counterpart in $(2+1)$-dimensions. We conclude with the case of the extended Green-Naghdi
equations.
\end{abstract}

\begin{highlights}

\item We revisit the derivation of the extended versions of the KdV, BBM and
Camassa-Holm equations in 1D, the extended cylindrical KdV equation in the
quasi-1D setting, the extended Kadomtsev-Petviashvili and its cylindrical
counterpart in 2D, as well the extended Green-Naghdi equations.

\item All the extended equations are valid one order beyond the usual weakly nonlinear,
long wave approximation and incorporate all appropriate dispersive and nonlinear terms.

\item The connection and applicability of the shallow water wave equations to other physical
contexts, including plasmas, nonlinear optics, Bose-Einstein condensation, solid mechanics
and others, are discussed.

\end{highlights}

\begin{keyword}
Shallow water waves \sep Nonlinear waves \sep Euler equations \sep Asymptotic expansions


\end{keyword}

\end{frontmatter}

\section{Introduction}

Water waves, forming and propagating on the surface of the ocean, are probably the most
commonly observed natural phenomenon.  While easy to observe, from a mathematical point of view
their modelling and analysis is very demanding as these waves feature significant differences from
waves propagating in other media. Indeed, water waves propagate on a free surface,
which is determined as part of the solution, so that the water wave equations form
a nonlinear free boundary value problem.

The study of water waves and their various ramifications remains central to fluid dynamics, and to
the dynamics of the oceans in particular, and plays--- in general--- a significant role in applied
mathematics and physics. The mathematical theory of water waves
\cite{stoker,whitham_book,johnson_book,lannes_book}, while being important on its own merit in
fluid mechanics, has provided the solid background and impetus for the development of the theory of
nonlinear dispersive waves in general, which has made a tremendous impact on numerous
disciplines \cite{Infeld,Dau}. Indeed, most of the fundamental ideas and results for nonlinear
dispersive waves, and particularly of dispersive shock waves ({\it alias} undular bores in
fluid mechanics), solitary waves and solitons, originated from the investigation
of water waves \cite{whitham_book,ist,ablowitz_book}.

Traditionally, the water wave problem \cite{stoker,whitham_book,johnson_book,lannes_book} consists
of studying the motion of the free fluid surface and the evolution of the velocity field of the
fluid under the following assumptions: the fluid is ideal, incompressible, irrotational and under
the influence of gravity and/or surface tension. As stated, this forms a nonlinear free surface
problem for which there are no general solutions. For this reason, various reductions of the full
water wave equations in various limits have been widely studied, including linear
(small amplitude) waves, weakly nonlinear waves, long waves and shallow water waves
\cite{whitham_book,lannes}. These limiting cases are not only attractive from a mathematical point
of view, but they also have wide applicability to the successful modelling of waves in nature
\cite{baines,grimshawbook,crighton,gen_book2,gen_book1,elreview}. In this work, we shall construct
equations which govern waves on the free surface of water, especially in the shallow water regime
of weakly nonlinear long waves. This limit 
results in ubiquitous equations, such as the Korteweg-de Vries (KdV)
equation \cite{whitham_book,ablowitz_book}
\begin{equation}
u_{t} + 6uu_x + u_{xxx} = 0,
 \label{e:kdv}
\end{equation}
where subscripts denote partial derivatives. The KdV equation is applicable to fields
outside of water wave theory, for instance plasma physics \cite{Infeld,berezin,jeffrey},
mechanical and electrical lattices \cite{Dau}, as well as in physical contexts where
the defocusing nonlinear Schr{\"o}dinger (NLS) equation plays a key role, e.g.,
nonlinear optics \cite{ofy}, nematic liquid crystals \cite{nematickdv} and atomic
Bose-Einstein condensates (BECs) \cite{huve,djf}. In these latter contexts, the fact that the
defocusing NLS can be asymptotically reduced to the KdV equation \cite{zakh}
(and, in two dimensions (2D), to the Kadomtsev-Petviashvili equation \cite{kuz2}--- see below)
allows for an effective description of dark NLS solitons in terms of KdV solitons.
This, in turn, enables the study of solitary waves of perturbed NLS models
that have the form of KdV solitons (see, e.g., Ref.~\cite{djfth} and
references therein). It is also important to mention that the KdV equation is the generic
nonlinear dispersive wave equation which is exactly integrable in a Hamiltonian sense,
and leads to the field of inverse scattering transform and soliton solutions
\cite{ist,whitham_book}.

The derivation of weakly nonlinear dispersive wave equations, such as the KdV equation and the
Boussinesq system \cite{whitham_book}--- which is the bidirectional equivalent of
the KdV equation--- is based on an asymptotic expansion of the water wave equations in the small
parameters of the wave amplitude to depth ratio and the depth to wavelength ratio. Balancing these
two small parameters results in these equations \cite{whitham_book}.
Truncating the KdV asymptotic expansion at the order of the small amplitude squared
results in the standard KdV equation. However, it has been found that, for many applications,
higher order terms in this asymptotic expansion are needed to adequately model physical waves.
The inclusion of all the next order terms
in the KdV asymptotic expansion results in the extended KdV (eKdV) equation \cite{ekdv}
\begin{equation}
u_{t} + 6uu_x + u_{xxx}
+ c_{1}u^{2}u_{x} + c_{2} u u_{xxx}
+ c_{3} u_{x} u_{xx}
+ c_{4} u_{xxxxx} = 0,
 \label{e:ekdv}
\end{equation}
where $c_{1} = -\alpha$, $c_{2} = 23/6 \alpha$, $c_{3} = 5/3 \alpha$ and $c_{4} = 19/60 \alpha$ for
the water wave eKdV equation, where $\alpha$ is the amplitude to depth ratio \cite{ekdv}.

The eKdV equation has been used to study solitary wave interaction \cite{soliton1,soliton2}, undular
bores \cite{hkdvbore}, the resonant flow of a fluid over topography \cite{reshkdv}, as well as
ion-acoustic waves in plasmas \cite{kodamatan}, and, more recently, dark solitons in weakly
nonlocal nonlinear media \cite{koutso1}. In particular, the inclusion of the higher order
terms of the eKdV equation was found to be vital to model and obtain agreement with experimental results on undular
bores in solid mechanics, with the KdV equation found to be an inadequate model \cite{karima}. Of relevance to the present review, the equations
governing nonlinear optical beams in dye doped nematic liquid crystals, which have a defocusing
response (i.e., the refractive index decreases with the optical intensity) can be reduced to the
eKdV equation \cite{nemgennady,saleh}. The higher order corrections present in the eKdV equation
are vital for the correct description of beam evolution, in particular for dispersive shock waves
as these are resonant.  The dispersion relation of the KdV equation is convex, so that resonance between an undular bore and dispersive radiation is not possible, while the higher order terms of the eKdV equation result in a non-convex linear dispersion relation, so that resonance
between an undular bore and linear dispersive waves is possible \cite{nemgennady,saleh}.

A particular case of the eKdV is the Kawahara equation \cite{kaw}, for which the
only higher order correction is a linear 5th-order dispersive term, i.e., $c_{1}=c_{2}=c_{3}=0$.
The Kawahara equation arises through the inclusion of capillary effects in the limit of the Bond
number being near $1/3$ \cite{patkdv}. It is important to point out that the inclusion of higher
order terms, and in particular of 5th-order dispersion, which can lead to a
non-convex linear dispersion relation, is vital for new effects not encompassed by the KdV
equation. Such effects include resonant solitary waves \cite{reskdv} and resonant
undular bores \cite{patkdv}, for which the solitary wave or waves of the undular bore are
in resonance with dispersive radiation. Solitary wave and undular bore resonance is strong
for the Kawahara equation \cite{reskdv,patkdv} and the resonant wavetrain can destroy the
classic undular bore structure, leaving just the resonant wavetrain \cite{patkdv}.
While the eKdV equation contains 5th-order dispersion, the strength of the
resonance is highly dependent on the relationship between the coefficients of the higher
order terms \cite{hkdvbore}. For the water wave coefficients, the amplitude of the resonant
wavetrain is very small \cite{prf}.

Another special case of the eKdV equation is the Gardner equation for which higher order
nonlinearity dominates over higher order dispersion, so that the only non-zero higher order
coefficient in the eKdV equation~(\ref{e:ekdv}) is $c_{1}$.
The Gardner equation is integrable, as for the KdV equation, and first arose in the derivation
of the infinite number of conservation laws for the KdV equation \cite{muira}. Since then
it has been widely studied and used to model water waves, for example to study large amplitude
surface and internal waves \cite{grimshawbook,helfrich,yury}, large amplitude undular bores
\cite{elgardner1} and resonant flow over topography \cite{ekdv,melville,elgardner2}. Furthermore,
the Gardner equation has been used in studies in plasma physics \cite{Infeld}, while it has
recently been proposed as a model describing internal, bright and dark, rogue waves in
three layer fluids \cite{Akhm}.

In addition to these one dimensional (1D) models, equations for quasi 1D propagation, as, e.g.,
in the case of radially symmetric wave structures, have been derived and analyzed in
the water waves context, as well as in other branches of physics. In the quasi-1D setting, a
pertinent model is the cylindrical KdV (cKdV) equation, which was derived in
Ref.~\cite{johnson2} for water waves, and later was used to describe cylindrical
solitary waves in plasmas \cite{maxon}, in nonlinear optical media with a local
\cite{rdsofy,bamdjf} or a nonlocal \cite{rdsnl1,rdsnl2} nonlinearity, and in
atomic BECs \cite{xue}. Additionally, higher order variants of the circular Korteweg-de Vries (cKdV) equation have also
been proposed and studied in the contexts of water waves \cite{guo} and plasma physics
\cite{mannan}.

On the other hand, a key model in the 2D setting is the $(2+1)$
dimensional equivalent of the KdV equation, the Kadomtsev-Petviashvili (KP) equation
\cite{Kadomtsev}, which incorporates weak lateral dispersion. Various types of generalized KP
equations, bearing various types of nonlinearity or incorporating 5th order
dispersion (from the dispersion relation of the water wave equations), have been studied. In
particular, theoretical results concerning the local well-posedness of higher order KP equations
\cite{kp1}, as well as the existence and non-existence of localized solitary wave solutions of
these generalized KP systems \cite{kp_book,kp2}, have been reported; see the survey~\cite{kp5}.
Importantly, much like the KdV equation, the KP equation and its variants appear in many contexts--- beyond water waves--- as an effective model for the study of the transverse dynamics
of 1D solitons in a 2D setting, for the description of weakly localized 2D solitons,
so-called ``lumps'' \cite{ist}, as well as for studies of soliton interactions.
Thus, the KP equation has been used in the study of the transverse instability of
dark NLS solitons \cite{kuz2} to predict the existence of 2D solitons in plasmas
\cite{Infeld}, optical media \cite{bullet}, atomic BECs \cite{hulump} and exciton-polariton
superfluids \cite{exci}, as well as to effectively describe soliton interactions in nonlocal
nonlinear media \cite{nemkp2}.

Furthermore, higher order KP equations, as well versions in
cylindrical coordinates--- first derived for water waves \cite{johnson2} and then
used extensively in plasma physics \cite{Infeld}--- have also appeared in many other contexts.
Indeed, in higher dimensions, the defocusing nematic equations describing nonlinear optical beam propagation in nematic liquid crystals can be
reduced to the KP equation \cite{nemkp1,nemkp2,nemkp3} and the cKdV equation \cite{nemckdv1}.
While the higher order corrections to these equations have not, as yet, been derived, these are anticipated to
be of the same form as the extended KP (eKP) and extended cylindrical KdV (ecKdV) equations of the
present review.  These higher order equations are necessary to model $(2+1)$ dimensional
dispersive shock waves in nematic liquid crystals.
Similar equations apply to optical beams in thermal nonlinear media \cite{kuz,barsi,trilloconti}, such as
lead glasses \cite{lead,segev,segev2} and liquids \cite{trilloconti}, and some photorefractive crystals \cite{barsi,fleischer,photo,conti_enc}.
In this respect, notice that thermal optical media are usually defocusing and so support optical undular bores.

The aim of this work is the derivation and systematisation of various extended,
dispersive shallow water wave equations arising
as asymptotic approximations from the water wave equations.
As is evident from the above discussion, the theme of extended shallow water
models is not only important for the water wave problem, but also for
many other physically important applications. The fact that water wave models may appear generically
as asymptotic reductions of other nonlinear evolution equations appearing, e.g., in
plasmas, optical systems, BECs, solid mechanics, etc, makes the methodology for
their systematic derivation particularly relevant and important.
This is also especially so because the derivation of the higher order corrections
to other nonlinear dispersive wave equations, e.g., the NLS equation, is similar to the derivations of the present review \cite{haskod}.

Furthermore, since our focus is
on extended shallow water wave equations, novel effects--- that can not be captured
by their ``non-extended'' counterparts--- are predicted to occur. Such effects are
pertinent to the evolution of steeper waves with shorter wavelengths, or pulse-widths
as in the case of pulse propagation in nonlinear optical fibres \cite{haskod,kivshar_book},
for which higher order corrections have been found to be particularly important. In addition, from a
mathematical point of view, the study of extended water wave models is
relevant to the theory of integrable systems. Indeed, the eKdV equation, for instance,
stems naturally from the underlying Hamiltonian systems \cite{menyuk,menyukchen},
and is related to the first higher-order equation in the KdV hierarchy \cite{Newell}.
In addition, the eKdV equation has been proved to be an important model in the context of
asymptotic integrability of weakly dispersive nonlinear wave equations \cite{fokasprl,kodamaint}.
It is thus anticipated that the derivation and presentation of extended shallow water
wave equations will be relevant to--- and will inspire new studies in--- a variety of physical
contexts and applied mathematics.

The organization of our presentation is as follows.
Firstly, in Section~2, we present the Euler equations and discuss the asymptotic
regimes for which the shallow water wave equations are relevant.
In Section~3, we revisit the derivation of the extended KdV
equation~\cite{ekdv}, and then derive other extended models in $(1+1)$ dimensions,
namely the extended Benjamin-Bona-Mahony (BBM)
and Camassa-Holm (CH) equations.
In Section~4 we consider a quasi-1D setting for shallow water waves of radial
symmetry, and derive the extended circular KdV
(cKdV) equation. In Section~5, we deal
with waves in $(2+1)$ dimensions and present the derivation of extended KP equations
in both Cartesian and cylindrical coordinates.
Furthermore, for completeness, in Section~6, we revisit the derivation of the
extended Green-Naghdi (GN) equations~\cite{matsuno2} in both $(1+1)$ and $(2+1)$ dimensions.
Finally, in Section~7, we summarize our conclusions.


\section{Mathematical formulation and asymptotic regimes}

We consider waves on the surface of an incompressible, inviscid fluid of undisturbed depth $h$.
The bottom boundary is taken to be flat. The $\tilde{z}$ direction is taken vertically upwards,
opposite to the direction of gravity, and the $(\tilde{x},\tilde{y})$ directions are horizontal.
Typical wavelengths in the $\tilde{x}$ and $\tilde{y}$ directions are $\lambda_{x}$ and
$\lambda_{y}$, respectively, and a typical wave amplitude is $a$.  For simplicity, the water wave
equations are transformed to non-dimensional form. The space variables
$(\tilde{x},\tilde{y},\tilde{z})$ are non-dimensionalised in the horizontal directions using the
typical wavelengths, $\tilde{x} = \lambda_{x} x$, $\tilde{y} = \lambda_{y}y$, and in the vertical
direction using the depth $h$, $\tilde{z} = hz$.  The velocity potential $\tilde{\phi}$ is made
non-dimensional based on the typical value $\lambda_{x} ga/c_{0}$, $\tilde{\phi} = (\lambda_{x}
ga/c_{0}) \phi$, where $g$ is the acceleration due to gravity and $c_{0} = \sqrt{gh}$ is the linear
long wave speed.  Time $\tilde{t}$ is non-dimensionalised on the typical value $\lambda_{x}/c_{0}$,
$\tilde{t} = (\lambda_{x}/c_{0})t$.  Finally, the surface displacement $\tilde{\eta}$ is
non-dimensionalised by the typical wave amplitude $a$, $\tilde{\eta} = a\eta$. The bulk motion of
the fluid is governed by Laplace's equation for the velocity potential $\phi$ \cite{whitham_book}
\begin{equation}
  \mu^2(\phi_{xx} + \delta^2\phi_{yy}) + \phi_{zz} = 0, \quad -1 < z < \varepsilon\eta.
  \label{e:laplace}
\end{equation}
This is solved together with the impenetrable boundary condition at the fluid bottom,
\begin{equation}
  {\phi _z} = 0,\quad z =  - 1,
\label{e:bottom}
\end{equation}
and the dynamic and kinematic boundary conditions at the fluid surface $z = \varepsilon \eta$,
respectively,
\begin{eqnarray}
  \phi_t + \frac{1}{2} \varepsilon \left[\phi^2_x + \delta^2\phi^2_y +
  \frac{1}{\mu^2}\phi^2_z\right] + \eta = 0, & &
      z=\varepsilon\eta,
      \label{e:dynamic} \\
  \mu^2\left[\eta_t + \varepsilon\left(\phi_x\eta_x +
          \delta^2\phi_y\eta_y\right)\right] = \phi_z, & &  z=\varepsilon\eta.
          \label{e:kinematic}
\end{eqnarray}
The dimensionless parameters appearing in these equations are $\varepsilon=a/h$, which is a measure
of nonlinearity, $\delta=\lambda_x/\lambda_y$ which measures the ratio of the wavelengths in the
$x$- and $y$- directions, and $\mu=h/\lambda_x$, which measures the strength of dispersion.
In the present work we consider weakly nonlinear long waves such that the wavelength is much
greater than the water depth, i.e., $\mu \ll 1$, and the wave amplitude is much less than the fluid
depth, so that $\varepsilon \ll 1$.
Our asymptotic analysis of the water wave equations
exploits these small parameters,
which has proven to be a powerful tool for deriving shallow water wave models of great utility,
see, e.g., Refs.~\cite{johnson_book,johnson4}. Other formulations \cite{fokasww} that may also include varying topography and
more general boundary conditions will be studied in another communication.

\begin{figure}[tbp]
\centering
\includegraphics[width=0.8\linewidth]{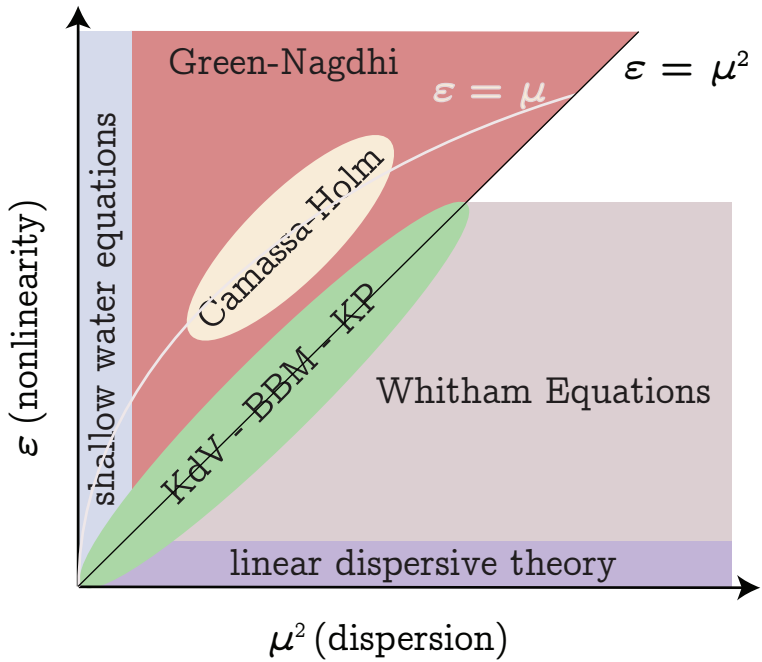}
\caption{(Color online) Schematic representation of the domains of validity of the various approximations to the water wave
equations considered in this review, in terms of the size
of the characteristic nonlinearity and dispersion parameters, $\varepsilon$ and $\mu$, respectively.}
\label{fig}
\end{figure}

To help put the many approximations to the water wave equations derived in this work in context, the domains of
validity of these various approximate equations are illustrated in a schematic fashion in
Fig.~\ref{fig} in the dispersion, measured by $\mu$, and nonlinearity, measured by $\varepsilon$,
plane.

The focus of this review is nonlinear, dispersive asymptotic reductions of the water wave
equations. The limit of no dispersion results in the shallow water equations \cite{whitham_book}
and the linear amplitude limit results in classical linear dispersive theory, which are not dealt
with in the current review.  These two limits are extensively dealt with in standard texts, see \cite{whitham_book}, for instance.
An extension of KdV-type equations
to include full dispersion is the Whitham equation \cite{whitham_book,craigetal05} and its extensions, as
discussed in Sections~\ref{s:ekdv} and \ref{s:ch}. The inclusion of full dispersion results
in solutions of these equations having effects, such as peaking and breaking, not present
in KdV-type approximations.

\section{(1+1) dimensional equations}
\label{s:1d}

\subsection{The extended Korteweg-de Vries equation}
\label{s:ekdv}

The eKdV equation will be rederived here {from the water wave equations} in a slightly different manner to
that of \cite{ekdv} in order to motivate the derivation of the other higher order weakly nonlinear, weakly
dispersive equations of this work and to put them into context.
This same procedure was later used
to further extend the theory, which goes beyond the KdV equation \cite{karc}.
KdV-type approximations lie around the line $\varepsilon=\mu^{2}$ in Figure \ref{fig}.
To derive the KdV equation from the above Euler system (\ref{e:laplace})--(\ref{e:kinematic}) consider the case $\delta=0$, so that waves
propagate in one spatial dimension and obey the reduced Euler system
\begin{subequations}
\begin{gather}
  {\phi _{zz}} + \varepsilon {\phi _{xx}} = 0, \quad  - 1 < z < \varepsilon \eta,
  \label{laplace1d}
  \\
  \frac{{\partial \phi }}{{\partial z}} = 0,\quad z = - 1,
  \label{bottom1d}
  \\
  {\phi _t} + \frac{1}{2}\varepsilon \left( {\phi _x^2 + \frac{1}{\varepsilon }\phi _z^2} \right)
  + \eta  = 0, \quad z = \varepsilon \eta,
\label{dynamic1d}
  \\
  \varepsilon \left( {{\eta _t} + \varepsilon {\phi _x}{\eta _x}} \right) = {\phi _z},
  \quad z = \varepsilon \eta.
  \label{kinematic1d}
\end{gather}
\label{euler1d}
\end{subequations}
We first solve Laplace's equation with the bottom boundary condition and then substitute into the
remaining free surface boundary conditions. As such, we expand the velocity potential $\phi$ as
\begin{equation}
\phi  = {\phi _0} + \varepsilon {\phi _1}
+ {\varepsilon ^2}{\phi _2} + {\varepsilon ^3}{\phi _3} +  \cdots,
\label{e:phiexpand}
\end{equation}
and then substitute this into Laplace's equation (\ref{laplace1d}), which results in
\[
{\phi _{0zz}} + \varepsilon ({\phi _{1zz}} + {\phi _{0xx}}) + {\varepsilon ^2}({\phi _{2zz}}
+ {\phi _{1xx}}) + {\varepsilon ^3}({\phi _{3zz}} + {\phi _{2xx}}) +  \cdots  = 0.
\]
After applying the bottom boundary condition (\ref{bottom1d}), we find
\begin{gather*}
  {\phi _0}(x,z,t) = A(x,t),\quad {\phi _1}(x,z,t) =  - \frac{{{{(z + 1)}^2}}}{2}{A_{xx}}, \\
  {\phi _2}(x,z,t) = \frac{{{{(z + 1)}^4}}}{{24}}{A_{xxxx}},\quad {\phi _3}(x,z,t)
  = -\frac{{{{(z + 1)}^6}}}{{720}}{A_{xxxxxx}}.
\end{gather*}
Finally, substituting for the potential $\phi$ into the dynamic (\ref{dynamic1d}) and kinematic
(\ref{kinematic1d}) boundary conditions, we find that up to $O(\varepsilon^2)$
\begin{gather}
\eta  + {A_t} + \varepsilon \left[ {\frac{1}{2}A_x^2 - \frac{1}{2}{A_{xxt}}} \right] +
{\varepsilon ^2}\left[ {\frac{1}{2}A_{xx}^2 - \eta {A_{xxt}} - \frac{1}{2}{A_x}{A_{xxx}}
+ \frac{1}{{24}}{A_{xxxxt}}} \right] = 0,
\label{t1} \\
{\eta _t} + {A_{xx}} + \varepsilon \left[ {{{(\eta {A_x})}_x} - \frac{1}{6}{A_{xxxx}}} \right]
+ {\varepsilon ^2}\left[ { - \frac{1}{2}{{(\eta {A_{xxx}})}_x}
+ \frac{1}{{120}}{A_{xxxxxx}}} \right] = 0.
\label{t2}
\end{gather}
Differentiating Eq.~(\ref{t1}) with respect to $x$ and defining $w=A_x$,
Eqs.~(\ref{t1})--(\ref{t2}) can finally be cast into the form
\begin{gather}
{w_t} + {\eta _x} + \varepsilon \left[ {w{w_x} - \frac{1}{2}{w_{xxt}}} \right]
+ {\varepsilon ^2}\left[ { - {{(\eta {w_{xt}})}_x} + \frac{1}{2}{w_x}{w_{xx}}
- \frac{1}{2}w{w_{xxx}} + \frac{1}{{24}}{w_{xxxxt}}} \right] = 0,
\label{bous1}\\
{\eta _t} + {w_x} + \varepsilon \left[ {{{(\eta w)}_x} - \frac{1}{6}{w_{xxx}}} \right]
+ {\varepsilon ^2}\left[ { - \frac{1}{2}{{(\eta {w_{xx}})}_x}
+ \frac{1}{{120}}{w_{xxxxx}}} \right] = 0,
\label{bous2}
\end{gather}
up to $O(\varepsilon^{2})$.
The above equations, (\ref{bous1}) and (\ref{bous2}), constitute a higher order Boussinesq-type
system \cite{whitham_book}. In order for the two equations to be compatible, let us define
\[
w = \eta  + \varepsilon {w_1} + {\varepsilon ^2}{w_2} + O({\varepsilon ^3}),
\]
so that now the system (\ref{bous1})--(\ref{bous2}) is
\begin{align*}
{\eta _t} + {\eta _x} &+ \varepsilon \left[ {{w_{1t}} + \eta {\eta _x} - \frac{1}{2}{\eta _{xxt}}} \right]
\\&+ {\varepsilon ^2}\left[ {{w_{2t}} + {{(\eta {w_1})}_x} - \frac{1}{2}{w_{1xxt}} - {{(\eta {\eta _{xt}})}_x} +
\frac{1}{2}{\eta _x}{\eta _{xx}} - \frac{1}{2}\eta {\eta _{xxx}} + \frac{1}{{24}}{\eta _{xxxxt}}} \right] = 0,  \\
{\eta _t} + {\eta _x} &+ \varepsilon \left[ {{w_{1x}} + 2\eta {\eta _x}
- \frac{1}{6}{\eta _{xxx}}} \right]
\\
&+ {\varepsilon
^2}\left[ {{w_{2x}} + {{(\eta {w_1})}_x} - \frac{1}{6}{w_{1xxx}} - \frac{1}{2}{{(\eta {\eta _{xx}})}_x} +
\frac{1}{{120}}{\eta _{xxxxx}}} \right] = 0.
\end{align*}
Direct comparison of these two equations, and replacing the time derivatives as
$${\eta _t} = -{\eta _x} -\varepsilon \left( {\frac{3}{2}\eta {\eta _x}
+ \frac{1}{6}{\eta _{xxx}}} \right),$$
results in
\[
   w = \eta  + \varepsilon \left( - \frac{1}{4}{\eta ^2} + \frac{1}{3}{\eta _{xx}} \right)
   + {\varepsilon ^2}\left( \frac{1}{8}{\eta ^3} + \frac{3}{{16}}\eta _x^2 + \frac{1}{2}\eta {\eta _{xx}}
  + \frac{1}{{10}}{\eta _{xxxx}} \right)+O(\varepsilon^3).
\]
This finally leads to the extended KdV equation, incorporating higher order dispersive and
nonlinear terms at one order beyond the KdV approximation
\begin{equation}
{\eta _t} + {\eta _x} + \varepsilon \left( {\frac{3}{2}\eta {\eta _x} + \frac{1}{6}{\eta
_{xxx}}} \right) + {\varepsilon ^2}\left( { - \frac{3}{8}{\eta ^2}{\eta _x} +
\frac{{23}}{{24}}{\eta _x}{\eta _{xx}} + \frac{5}{{12}}\eta {\eta _{xxx}} + \frac{{19}}{{360}}{\eta
_{xxxxx}}} \right) = 0,
\label{ekdv}
\end{equation}
as found in \cite{ekdv} on a slight rescaling of the equation.
{This eKdV equation with these coefficients on the higher order $\epsilon^{2}$ terms is strictly applicable to water waves.  However, it is applicable in the more general form (\ref{e:ekdv}) to other weakly nonlinear, weakly dispersive waves, such as internal waves in a stratified fluid, with the coefficients $c_{i}$, $i=1,\ldots,4$, taking values applicable to the specific application.
The higher order terms in the eKdV equation (\ref{e:ekdv}) introduce new effects not present for the integrable KdV equation.  In particular, fifth order dispersion leads to resonance between solitary waves, undular bores and other structures and the eKdV equation is then not integrable \cite{patkdv}.}

\subsection{The extended Benjamin-Bona-Mahony equation}

Another scalar $(1+1)$-dimensional model associated with weakly nonlinear long wave reductions of the water wave equations and the bidirectional Boussinesq system is the Benjamin-Bona-Mahony (BBM) equation \cite{bona1}
\[
{\eta _t} + {\eta _x} + \varepsilon \left( {\frac{3}{2}\eta {\eta _x} -
\frac{1}{6}{\eta _{xxt}}} \right)=0.
\]
While this model was originally introduced for the study of undular bores \cite{peregrine} (cf.\ Ref.\ \cite{congy} for a
classification of solutions of the dispersive Riemann problem for the BBM equation), in principle,
numerical solutions of the BBM equation have improved stability over those of the KdV equation
under the unidirectional assumption in $(1+1)$ dimensions due to its bounded dispersion relation.
The issue of the improved
numerical stability of the BBM equation over the KdV equation has become less of an issue due to
the development of improved numerical methods for the KdV equation, see \cite{chan,tref}, for
instance. Note that, contrary to the infinite number of integrals of motion of the KdV
equation, the BBM equation possesses only three conservation laws and is not integrable,
so that it has received much less attention than the KdV equation. The properties of surface
water waves in a channel governed by the BBM equation have been discussed in Ref.~\cite{bona3}.

At leading order the KdV and BBM equations are closely related as they are asymptotically the same.
At $O(\varepsilon)$ the change $\eta_x=-\eta_t$ can be made in the dispersive term of the eKdV
equation~(\ref{ekdv}), since this relation holds at $O(1)$. In a similar fashion, the change
\[
{\eta _{xxx}} =  - {\eta _{xxt}} - \varepsilon \frac{\partial^2}{\partial x^2}{\left( {\frac{3}{2}\eta {\eta _x}
- \frac{1}{6}{\eta _{xxt}}} \right)},
\]
in the eKdV Eq.~(\ref{ekdv}) results in the higher order correction of the BBM equation
\begin{equation}
{\eta _t} + {\eta _x} + \varepsilon \left( {\frac{3}{2}\eta {\eta _x}
- \frac{1}{6}{\eta _{xxt}}} \right)
+ {\varepsilon ^2}\left( { - \frac{3}{8}{\eta ^2}{\eta _x} + \frac{5}{{24}}{\eta _x}{\eta _{xx}} -
\frac{1}{6}\eta {\eta _{xxt}} - \frac{1}{{40}}{\eta _{xxxxt}}} \right) = 0,
\label{ebbm}
\end{equation}
which is the extended BBM (eBBM) equation. Alternatively, the compatibility condition for
Eqs.~(\ref{bous1}) and (\ref{bous2}) in the derivation of the eKdV equation may
be constructed as
\[
  w = \eta  + \varepsilon \left( - \frac{1}{4}{\eta ^2} - \frac{1}{3}{\eta _{xt}} \right)
  + {\varepsilon ^2}\left( \frac{1}{8}{\eta ^3} - \frac{5}{{16}}\eta _x^2 -
  \frac{2}{{45}}{\eta _{xxxt}} \right)+O(\varepsilon^3).
\]
Then the change
\[
{\eta _{xtt}} =  - {\eta _{xxt}} - \varepsilon
\frac{\partial^{2}}{\partial x \partial t}\left( {\frac{3}{2}\eta {\eta _x}
- \frac{1}{6}{\eta _{xxt}}} \right),
\]
also results in the eBBM equation~(\ref{ebbm}).

\subsection{The extended Camassa-Holm equation}
\label{s:ch}

Both the KdV and BBM equations are globally well posed and, therefore, these models are not able to
capture wave breaking (recall that, in both models, dispersion exactly balances nonlinearity, $\epsilon = \mu^{2}$).
Nevertheless, there exist other shallow water wave equations which avoid this drawback by including
either short wave effects or stronger nonlinearity. One such model is the so-called Whitham
equation \cite{whitham_book,craigetal05,whith1}, which includes short wave effects by replacing the dispersive
term $u_{xxx}$ in the KdV equation by a Fourier integral whose kernel is the full
water wave dispersion relation, so that the Whitham equation applies in a region to the right of the
KdV region in Figure~\ref{fig} with higher $\mu$.
It has been shown that the Whitham equation possesses solutions
showing both peaking and breaking, as for real water waves \cite{whitham_book,whith1}.  While
Whitham derived this equation on an ad-hoc basis, it has subsequently been shown that it can be
rigorously derived from the Hamiltonian formulation of the water wave equations \cite{craigetal05}.
Another model, characterized by a stronger nonlinearity, so that it lies above the KdV region in
Figure~\ref{fig}, is the Camassa-Holm (CH) equation \cite{ch}
\begin{equation}
{u_T} + \frac{{10}}{{19}}{u_X} + \varepsilon \left( \frac{3}{2} u{u_X}
  - \frac{19}{60} u_{TXX} \right) + {\varepsilon ^2}\left( { - \frac{{19}}{{60}}{u_X}{u_{XX}}
  - \frac{{19}}{{120}}u{u_{XXX}}} \right)=0,
\end{equation}
where the variables $X$ and $T$ are defined in a travelling frame.
Interestingly, the CH equation was first discovered as a completely integrable equation
\cite{fokas}, while its relevance to the description of shallow water waves gained notice
much later \cite{ch}. As seen by its form, the CH equation captures stronger nonlinear effects
than the nonlinear dispersive KdV and BBM equations and can
indeed accommodate wave breaking phenomena \cite{constantin}. Furthermore,
the nonlinear dispersive terms $u_{X}u_{XX}$ and $u u_{XXX}$ in this equation are the same as the
higher order nonlinear dispersive terms in the eKdV equation (\ref{ekdv}).

Below, we shall show that the starting point for the derivation of the CH equation, and its
extended form, the eCH equation, is the KdV equation. Indeed, the CH equation cannot be derived
directly from the Euler system using the analysis of Section~\ref{s:ekdv} for the eKdV equation
\cite{johnson3,bhatt}. On the other hand, we note that, alternatively, the CH equation can be
derived by a variational approach in a Lagrangian formalism of the water wave equations
\cite{kruse}, or via the Green-Naghdi equations \cite{johnson3}.

Before proceeding with the derivation of the extended CH equation, it is useful to make some
remarks.
The integrable KdV equation emerges at first order in an asymptotic expansion for
unidirectional shallow water waves. However, at quadratic order, this asymptotic expansion
produces an entire family of shallow water wave equations that are asymptotically equivalent to
each other under a group of nonlinear, nonlocal, normal form transformations \cite{kodama} in
combination with the application of the Helmholtz operator. These transformations can be used to
present connections between shallow water waves, the integrable fifth-order KdV equation and a
generalization of the CH equation that contains an additional integrable case. The linear
dispersion relation for the full water wave equations for arbitrary depth and any equation in this
family agree to fifth order. The travelling wave solutions of the CH equation have been shown to
asymptotically agree with the exact solution of the fifth order KdV equation \cite{dullin2}.
The formal
analysis for parameter ranges consistent with the asymptotic derivation of the CH equation can be
found in Ref.~\cite{parker}, while it has been shown that the solutions of the higher order KdV
equation can be related to solutions of the CH equation \cite{dullin1}.

To derive the higher order, to $O(\varepsilon^3)$, correction to the KdV equation
we use the water wave equations, Eqs.~(\ref{euler1d}), and take into account the relevant
additional terms. Following the methodology of Section~\ref{s:ekdv}, we first derive the
higher order Boussinesq system to $O(\varepsilon^{2})$ using an expansion of the form
(\ref{e:phiexpand}), namely
\begin{align*}
{w_t} + {\eta _x} &+ \varepsilon \left[ {w{w_x} - \frac{1}{2}{w_{xxt}}} \right]
+ {\varepsilon ^2}\left[ { - {{(\eta {w_{xt}})}_x} + \frac{1}{2}{w_x}{w_{xx}}
- \frac{1}{2}w{w_{xxx}}
+ \frac{1}{{24}}{w_{xxxxt}}} \right] \\
& \mbox{} + {\varepsilon ^3}\left[ {w_x^2{\eta _x} - \eta {\eta _x}{w_{xt}}
+ \eta {w_x}{w_{xx}}- {\eta _x}w{w_{xx}}
- \frac{1}{2}{\eta ^2}{w_{xxt}} - \eta w{w_{xxx}} + \frac{1}{{12}}{w_{xx}}{w_{xxx}}} \right]
 \\
&\mbox{} + {\varepsilon ^3}\left[ {\frac{1}{6}{\eta _x}{w_{xxxt}} - \frac{1}{8}{w_x}{w_{xxxx}}
+ \frac{1}{6}\eta {w_{xxxxt}}
+ \frac{1}{{24}}w{w_{xxxxx}} - \frac{1}{{720}}{w_{xxxxxxt}}} \right] = 0, \\
{\eta _t}+ {w_x} &+ \varepsilon \left[ {{{(\eta w)}_x} - \frac{1}{6}{w_{xxx}}} \right]
+ {\varepsilon ^2}\left[ { - \frac{1}{2}{{(\eta {w_{xx}})}_x}
+ \frac{1}{{120}}{w_{xxxxx}}} \right] \\
   &+ {\varepsilon ^3}\left[ { - \eta {\eta _x}{w_{xx}} - \frac{1}{2}{\eta ^2}{w_{xxx}}
   + \frac{1}{{24}}{{(\eta {w_{xxxx}})}_x} - \frac{1}{{5040}}{w_{xxxxxxx}}} \right] = 0.
\end{align*}
To make these two equations compatible, we set
$$  w = \eta  + \varepsilon {w_1} + {\varepsilon^2}{w_2}
+ {\varepsilon ^3}{w_3}+O(\varepsilon^4),$$
with
\begin{align*}
 {w_1} &=  - \frac{1}{4}{\eta ^2} + \frac{1}{3}{\eta _{xx}}, \quad
  {w_2} = \frac{1}{8}{\eta ^3} + \frac{3}{{16}}\eta _x^2 + \frac{1}{2}\eta {\eta _{xx}}
  + \frac{1}{{10}}{\eta _{xxxx}}, \\
   {w_3} &= \frac{3}{{16}}{\partial_x^{-1}}(\eta _x^3) - \frac{5}{{64}}{\eta ^4}
  + \frac{3}{{32}}\eta \eta _x^2
  + \frac{1}{8}{\eta ^2}{\eta _{xx}} + \frac{{163}}{{360}}\eta _{xx}^2
  + \frac{{1091}}{{1440}}{\eta _x}{\eta _{xxx}}\\
  &+ \frac{7}{{20}}\eta {\eta _{xxxx}} + \frac{{61}}{{1890}}{\eta _{xxxxxx}},
\end{align*}
where $$\displaystyle \partial^{-1}_X(u)=\int_{-\infty}^{X} u(X',T)\, dX'.$$
In this manner, we derive the higher order, $O(\varepsilon^3)$, KdV equation
\begin{gather}
{\eta _t} + {\eta _x} + \varepsilon \left( {\frac{3}{2}\eta {\eta _x}
+ \frac{1}{6}{\eta _{xxx}}} \right)
+ {\varepsilon ^2}\left( { - \frac{3}{8}{\eta ^2}{\eta _x} + \frac{5}{{12}}\eta{\eta _{xxx}}
+ \frac{{23}}{{24}}{\eta _x}{\eta _{xx}}
+ \frac{{19}}{{360}}{\eta _{xxxxx}}} \right) \nonumber \\
+ {\varepsilon ^3}\left( \frac{3}{{16}}{\eta ^3}{\eta _x}
+ \frac{{23}}{{16}}\eta {\eta _x}{\eta _{xx}}
+ \frac{5}{{16}}{\eta ^2}{\eta _{xxx}} + \frac{{19}}{{32}}\eta_x^3
+ \frac{{1079}}{{1440}}{\eta _x}{\eta _{xxxx}}\right. \nonumber\\
\left.
+ \frac{{317}}{{288}}{\eta _{xx}}{\eta _{xxx}} + \frac{{19}}{{80}}\eta{\eta _{xxxxx}}
+ \frac{{55}}{{3024}}{\eta _{xxxxxxx}} \right) = 0.
\label{kdv3}
\end{gather}
To transform the above $O(\varepsilon^3)$ version of the KdV equation into a form which can be
rescaled to the CH equation, first we introduce the Galilean transformation
\begin{equation}
\label{gal}
X = x - \frac{9}{19} t,\quad  T =  t,
\end{equation}
and perform the Kodama transformation \cite{kodama}
\[
\eta (X,T) = u + \varepsilon \left[ {\frac{7}{{20}}{u^2} + \frac{1}{{30}}{u_{XX}}
- \frac{1}{5}{u_X}\partial _X^{ - 1}(u)} \right],
\]
to express the resulting equation in terms of the new dependent variable $u=u(X,T)$. To this end,
applying the Helmholtz operator
$$\displaystyle{\cal H} = 1 - \varepsilon\frac{19}{60} \partial_X^2$$
to the resulting equation, we find that the third order KdV Eq.~(\ref{kdv3})
becomes the extended CH (eCH) equation
\begin{align}
  {u_T} &\mbox{} + \frac{{10}}{{19}}{u_X} + \varepsilon \left[ \frac{3}{2} u{u_X}
  - \frac{19}{60} u_{TXX} \right]
+ {\varepsilon ^2}\left[ { - \frac{{19}}{{60}}{u_X}{u_{XX}} - \frac{{19}}{{120}}uu_{XXX}} \right]
\nonumber \\
&\mbox{} + {\varepsilon ^3}\left[ {\frac{{223}}{{151200}}{u_{XXXXXXX}}
- \frac{{3}}{{100}}{u_{XX}}\left( {{u^2} - 2{u_X}\partial _X^{ - 1}\left( u \right)} \right)\partial
_X^{ - 1}\left( u
\right)} \right] \nonumber \\
&\mbox{} + \frac{{{\varepsilon ^3}}}{{2400}}\left[ {976 u{u_X}{u_{XX}}
+ 680{u^2}{u_{XXX}}} { + 2765 u_X^3 - 48u{u_{XXXX}}\partial _X^{ - 1}(u) + 48 u_{XX}^2\partial _X^{
- 1}\left( u \right)}
\right] \nonumber \\
   &\mbox{} + \frac{{{\varepsilon ^3}}}{{3600}}\left[ {903 {u_X}{u_{XXXX}} + 316 u{u_{XXXXX}} +
   305{u_{XX}}{u_{XXX}}} \right] = 0.
\label{ech}
\end{align}
%

\section{The cylindrical KdV equation}

The above derivations of the extended KdV, BBM and CH equations have been for $(1+1)$ dimensional
waves, that is waves propagating in one horizontal direction only. It is clear that most water
waves in nature are not one dimensional (1D), so the extended equations of the previous Sections
will now be extended to $(2+1)$ dimensions.
We first consider the quasi-1D setting and consider the water wave equations
(\ref{e:laplace})--(\ref{e:kinematic}) with radial symmetry. In this case, the water wave equations
can be expressed in cylindrical polar coordinates as
\begin{subequations}
\begin{gather}
  {\phi _{zz}} + \varepsilon \left( {{\phi _{rr}} + \frac{1}{r}{\phi _r}} \right) = 0,
  \label{ceuler1} \\
  {\phi _z} = 0,\quad z =  - 1,
  \label{ceuler2} \\
  {\phi _t} + \frac{\varepsilon }{2}\left( {\phi _r^2 + \frac{1}{\varepsilon }\phi _z^2} \right)
  + \eta  = 0,\quad z = \varepsilon \eta,
  \label{ceuler3} \\
  \varepsilon({\eta _t} + \varepsilon {\phi _r}{\eta _r}) = \phi _z,
  \quad z = \varepsilon \eta.
  \label{ceuler4}
\end{gather}
\label{ceuler}
\end{subequations}
From this system, Johnson \cite{johnson2} derived (see also \cite{johnson_book}) the so-called
cylindrical KdV (cKdV) equation
\begin{equation}
 {H_T} + \frac{3}{2}H{H_R} + \frac{1}{6}{H_{RRR}} + \frac{1}{{2T}}H =0.
 \label{e:ckdv}
\end{equation}
Here, $H$ is the rescaled wave amplitude of the free boundary, which depends on the independent
slow variables $R$ and $T$, defined in a proper travelling frame (see below). It should be noted
that the cKdV equation was presented in an earlier work by the same author \cite{johnson1}, where
it was also shown that the Inverse Scattering Transform for the cKdV can be obtained directly from
that for the Kadomtsev-Perviashvili (KP) equation. Since this work, relevant cKdV models have been
derived with the addition of surface tension \cite{guo}, while experiments were performed which
compared axisymmetric free surface solitary waves with theoretical and numerical solutions of the
cKdV equation, with good agreement found \cite{soliton3}.

Following the above mentioned previous work \cite{johnson_book,johnson2}, we introduce
the new variables
\[
R = \varepsilon (r - t), \quad T = {\varepsilon ^4}t,
\quad \phi  = \varepsilon \Phi, \quad \eta  = {\varepsilon ^2}H,
\]
and expand the scaled velocity potential $\Phi$ asymptotically as
\[
\Phi=\Phi_0+\varepsilon^3\Phi_1+\varepsilon^6\Phi_2+\varepsilon^9\Phi_3+\cdots.
\]
Laplace's equation (\ref{ceuler1}) then gives
\begin{align*}
  (R + T/{\varepsilon ^3}){\Phi _{0zz}} &+ T({\Phi _{1zz}} + {\Phi _{0RR}})
   + {\varepsilon ^3}\left[ {R{\Phi _{1zz}} + T{\Phi _{2zz}} + {{(R{\Phi _{0R}})}_R} + T{\Phi
   _{1RR}}} \right] \\
   &\mbox{} + {\varepsilon ^6}\left[ {R{\Phi _{2zz}} + T{\Phi _{3zz}} + {{(R{\Phi _{1R}})}_R} + T{\Phi
   _{2RR}}} \right] =
   O({\varepsilon ^9}).
\end{align*}
Solving this equation for each order of $\varepsilon$, and applying the bottom condition
Eq.~(\ref{ceuler2}), yields
\begin{align*}
  {\Phi _0} &= A(R,T),\quad
  {\Phi _1} =  - \frac{{{{\left( {z + 1} \right)}^2}}}{2}{A_{RR}},\quad
  {\Phi _2} =  - \frac{{{{\left( {z + 1} \right)}^2}}}{{2T}}{A_R}
  + \frac{{{{\left( {z + 1} \right)}^4}}}{{24}}{A_{RRRR}},\nonumber \\
  {\Phi _3} &= \frac{{{{\left( {z + 1} \right)}^2}R}}{{2{T^2}}}{A_R}
  + \frac{{{{\left( {z + 1} \right)}^4}}}{{12T}}{A_{RRR}}
  - \frac{{{{\left( {z + 1} \right)}^6}}}{{720}}{A_{RRRRRR}},
\end{align*}
where any homogeneous solutions that arise in higher order terms are absorbed into the
leading-order term $\Phi_0$. These solutions are then substituted back into the dynamic,
Eq,~(\ref{ceuler3}), and kinematic, Eq.~(\ref{ceuler4}), boundary conditions. It should
be noted that, in this work, we keep terms up to $O(\varepsilon^6)$.

Differentiating the dynamic boundary condition (\ref{ceuler3}) with respect to $R$ produces
\begin{gather}
  {H_R} - {w_R} + {\varepsilon ^3}\left[ {{w_T} + w{w_R} + \frac{1}{2}{w_{RRR}}} \right]
  \nonumber \\
   \mbox{} + {\varepsilon ^6}\left[ {\frac{1}{{2T}}{w_{RR}} + {H_R}{w_{RR}} + \frac{1}{2}{w_R}{w_{RR}}
   - \frac{1}{2}{w_{RRT}} + H{w_{RRR}} - \frac{1}{2}w{w_{RRR}}
   - \frac{1}{{24}}{w_{RRRRR}}} \right] = 0,
   \label{ber}
\end{gather}
while the kinematic boundary condition Eq.~(\ref{ceuler4}) reads
\begin{gather}
   - {H_R} + {w_R} + {\varepsilon ^3}\left[ {\frac{1}{T}w + {H_T} + {{(Hw)}_R}
   - \frac{1}{6}{w_{RRR}}} \right] \nonumber \\
 \mbox{}  + {\varepsilon ^6}\left[ { - \frac{R}{{{T^2}}}w + \frac{1}{T}Hw - \frac{1}{{3T}}{w_{RR}}
   - \frac{1}{2}{{(H{w_{RR}})}_R} + \frac{1}{{120}}{w_{RRRRR}}} \right] = 0,
\label{kin}
\end{gather}
to $O(\varepsilon^{6})$, where we have introduced $w=A_R$. The above equations, (\ref{ber}) and (\ref{kin}), now need to be consistent. To enforce this we set
$$w=H+ {\varepsilon ^3}{w_1} + {\varepsilon
^6}{w_2}+O(\varepsilon^9).$$
Compatibility then gives $w_1$ and $w_2$ as
\begin{align*}
  {w_1} &=  - \frac{1}{4}{H^2} + \frac{1}{3}{H_{RR}} - \frac{1}{{2T}}(\partial _R^{ - 1}H),
  \\
  {w_2} &= \frac{1}{8}{H^3} + \frac{3}{{16}}H_R^2 + \frac{1}{2}H{H_{RR}}
  + \frac{1}{{10}}{H_{RRRR}}
   + \frac{1}{T}\left[ {\frac{1}{6}{H_R} - \frac{1}{{16}}(\partial _R^{ - 1}{H^2})} \right]\\
   &+ \frac{1}{{{T^2}}}\left[ {\frac{1}{2}(\partial _R^{ - 1}RH)
   + \frac{5}{8}(\partial _R^{ - 2}H)} \right].
\end{align*}
With this compatibility enforced either of Eqs.~(\ref{ber}) or (\ref{kin}) give the
extended cylindrical KdV (ecKdV) equation
\begin{gather}
  {H_T} + \frac{3}{2}H{H_R} + \frac{1}{6}{H_{RRR}} + \frac{1}{{2T}}H \nonumber \\
 \mbox{}   + {\varepsilon ^3}\left[ { - \frac{3}{8}{H^2}{H_R} + \frac{{23}}{{24}}{H_R}{H_{RR}}
   + \frac{5}{{12}}H{H_{RRR}} + \frac{{19}}{{360}}{H_{RRRRR}}} \right] \nonumber \\
   \mbox{} + \frac{{{\varepsilon ^3}}}{T}\left[ {\frac{3}{{16}}{H^2} + \frac{1}{4}{H_{RR}}
   - \frac{1}{2}{H_R}\partial _R^{ - 1}(H)} \right]
   + \frac{{{\varepsilon ^3}}}{{{T^2}}}\left[{ - \frac{R}{2}H
   + \frac{1}{8}\partial _R^{ - 1}(H)} \right] = 0,
\label{ckdv}
\end{gather}
where $$\displaystyle\partial_R^{-1}H=\int_{0}^{R}H(R',T)\,dR'.$$
{A restricted form of the ecKdV equation with only higher order fifth order dispersion, $H_{RRRRR}$, was derived by Huang and Dai \cite{huang}.}

What is of interest here is that if the terms involving $1/T$ and $1/T^2$ are
ignored, then the eKdV equation (\ref{ekdv}) is recovered. This is expected, since as time
increases the waves propagate in $R$, which means that their radii of curvature decrease with the
wavefronts become increasingly flat, so that the waves become $(1+1)$ dimensional. While the water
wave equations (\ref{ceuler}) have no dependence on the radial angle $\theta$, they still govern
the propagation of radially symmetric waves, which is not a 1D effect. Thus, the higher
dimensionality involved in radially symmetric waves introduces nonlocal terms, being encompassed by
the operator $\partial_R^{-1}$. This induced nonlocality in higher dimensions will become more
apparent in the case of the KP equation (see Section \ref{s:ekp}). Solutions of the 2D wave equation also show
such nonlocality--- see Section~7.4 of Ref.~\cite{whitham_book}.

We conclude this Section with an important remark on the connection between the ecKdV equation
(\ref{ckdv}) and its Cartesian counterpart. For the regular cKdV equation and integrable KdV equation case, there
exists a transformation that links one equation to the other \cite{johnson1,hirota}. There also
exists such a transformation which maps the ecKdV equation (\ref{ckdv}) to the form of a perturbed
KdV equation with non-constant coefficients. Indeed, upon defining
\begin{equation}
H = \frac{R}{{3T}} - \frac{1}{{2T}}u(\xi ,\tau )
- \frac{{4{\varepsilon ^3}}}{3}{\xi ^2}{\tau ^2}\log \tau ,\quad R
= \frac{\xi }{\tau },\quad T =  - \frac{1}{{2{\tau ^2}}},
\label{e:ckdvtrans}
\end{equation}
one may express the ecKdV equation (\ref{ckdv}) as
\begin{gather}
  {u_\tau } + \frac{3}{2}u{u_\xi } + \frac{1}{6}{u_{\xi \xi \xi }}
  + {\varepsilon ^3}\left[ { - \frac{\xi }{6}(11 + 24\log \tau )u
  - \frac{{{\xi ^2}}}{2}(1 + 4\log \tau ){u_\xi }
  - \frac{1}{6}\partial _\xi ^{ - 1}(u)} \right]
  \nonumber \\
\mbox{} + {\varepsilon ^3}\tau \left[ { - \frac{1}{8}{u^2} + \frac{1}{2}\xi u{u_\xi }
   - \frac{{41}}{{36}}{u_{\xi \xi }}
   - \frac{5}{{18}}\xi {u_{\xi \xi \xi }} + {u_\xi }\partial _\xi ^{ - 1}(u)} \right]
   \nonumber \\
\mbox{} + {\varepsilon ^3}{\tau ^2}\left[ { - \frac{3}{8}{u^2}{u_\xi }
   + \frac{{23}}{{24}}{u_\xi }{u_{\xi \xi }}
   + \frac{5}{{12}}u{u_{\xi \xi \xi }} + \frac{{19}}{{360}}{u_{\xi \xi \xi \xi \xi }}} \right] =
   O({\varepsilon ^6}).
   \label{ckdv.kdv}
\end{gather}
Note that here the $O(\varepsilon^3)$ correction to the original transformation
for $H$ in (\ref{e:ckdvtrans}) serves to cancel any inhomogeneous terms
produced up to that order.

\section{(2+1) dimensional equations}

\subsection{The extended Kadomtsev-Petviashvili equation}
\label{s:ekp}

Here, we derive the full extended KP equation{from the water wave equations, as originally derived by \cite{jia},} which incorporates all possible higher order
correction terms at the next order of approximation beyond the standard KP equation, as was done
for the KdV equation in Section \ref{s:ekdv}.   The water wave equations will again be approximated using an asymptotic
expansion. The parameter $\delta = \lambda_{x}/\lambda_{y}$, the ratio of the wavelengths in the
two horizontal directions, is taken as $\delta^2\mapsto\varepsilon\delta^2$ in Laplace's equation
(\ref{e:laplace}). While this parameter can be absorbed via a change in the coordinates,
we opt to keep it in the wave equations so as to act as a measure of the dimensionality
contribution. KP-type equations are valid in same the KdV-type region of Figure~\ref{fig}.

As for the previous $(1+1)$ dimensional equations we expand the velocity potential as in
Eq.~(\ref{e:phiexpand}), and substitute back into Laplace's equation (\ref{e:laplace}) to obtain
\begin{align*}
{\phi _{0zz}} + \varepsilon ({\phi _{1zz}} + {\phi _{0xx}} + {\delta ^2}{\phi _{0yy}})
&+ {\varepsilon ^2}({\phi _{2zz}} + {\phi _{1xx}} + {\delta ^2}{\phi _{1yy}})
+ {\varepsilon ^3}({\phi _{3zz}} + {\phi _{2xx}} + {\delta ^2}{\phi _{2yy}})
= O(\varepsilon^4).
\end{align*}
Solving this differential equation at each order of $\varepsilon$, and applying the bottom
boundary condition (\ref{e:bottom}), gives
\begin{gather*}
  {\phi _0}(x,y,z,t) = A(x,y,t),\quad
  {\phi _1}(x,y,z,t) =  - \frac{{{{(z + 1)}^2}}}{2}({A_{xx}} + {\delta ^2}{A_{yy}}),\\
  {\phi _2}(x,y,z,t) = \frac{{{{(z + 1)}^4}}}{{24}}({A_{xxxx}} + 2{\delta ^2}{A_{xxyy}}
  + {\delta ^4}{A_{yyyy}}), \\
  {\phi _3}(x,y,z,t) = -\frac{{{{(z + 1)}^6}}}{{720}}({A_{xxxxxx}}
  + 3{\delta ^2}{A_{xxxxyy}} + 3{\delta ^4}{A_{xxyyyy}} + {\delta ^6}{A_{yyyyyy}}).
\end{gather*}
Next, differentiating the dynamic boundary condition  (\ref{e:dynamic}) with respect to $x$,
substituting the above components of the velocity potential and introducing $w=A_x$, casts the
dynamic (\ref{e:dynamic}) and kinematic (\ref{e:kinematic}) boundary conditions into the form
\begin{gather*}
  {w_t} + {\eta _x} + \varepsilon \left( {w{w_x} - \frac{1}{2}{w_{xxt}}} \right) +  \\
  {\varepsilon ^2}\left[{\delta ^2}{w_y}\partial _x^{ - 1}({w_y}) - \frac{1}{2}{\delta ^2}{w_{yyt}}
  - {\eta _x}{w_{xt}} + \frac{1}{2}{w_x}{w_{xx}} - \eta {w_{xxt}} - \frac{1}{2}w{w_{xxx}}
  + \frac{1}{{24}}{w_{xxxxt}}\right] = 0,
\\
  {\eta _t} + {w_x} + \varepsilon \left[{\delta ^2}\partial _x^{ - 1}({w_{yy}}) + {(\eta w)_x}
  - \frac{1}{6}{w_{xxx}}\right] \\
\mbox{} + {\varepsilon ^2}\left[{\delta ^2}\eta \partial _x^{ - 1}({w_{yy}})
   + {\delta ^2}{\eta _y}\partial _x^{ - 1}({w_y})
   - \frac{1}{3}{\delta ^2}{w_{xyy}} - \frac{1}{2}{(\eta {w_{xx}})_x}
   + \frac{1}{{120}}{w_{xxxxx}}\right] = 0.
\end{gather*}
where, as before, $\displaystyle \partial^{-1}_x (u)=\int_{-\infty}^{x}u(x',y,t)\, dx'$.
In order for these two equations to be consistent, we again set
$$ w = \eta + \varepsilon ({w_1} + {\delta
^2}{w_{12}}) + {\varepsilon ^2}({w_2} + {\delta ^2}{w_{21}})+O(\varepsilon^3),$$
where we isolate different corrections to emphasize the role of dimensionality. Then,
\begin{align*}
  {w_1} &=  - \frac{1}{4}{\eta ^2} + \frac{1}{3}{\eta _{xx}},
  \quad
  w_{12} =- \frac{{{1}}}{2}\partial _x^{ - 2}({\eta _{yy}}), \quad
  {w_2} = \frac{1}{8}{\eta ^3} + \frac{3}{{16}}\eta _x^2 + \frac{1}{2}\eta {\eta _{xx}}
  + \frac{1}{{10}}{\eta _{xxxx}}, \\
  {w_{21}} &= \frac{1}{6}{\eta _{yy}}
  - \frac{3}{8}\partial _x^{ - 1}(\eta \partial _x^{ - 1}({\eta _{yy}}))
  + \frac{5}{8}\partial _x^{ - 2}(\eta {\eta _{yy}} + \eta _y^2)
  + \frac{{3{\delta ^2}}}{8}\partial _x^{ - 4}({\eta _{yyyy}}).
\end{align*}
Note that $w_1$ and $w_2$ have already been derived in Ref.~\cite{ekdv}, in which the derivation of
the extended KdV equation was presented.  In this manner, under unidirectional propagation, the
extended KP (eKP) equation is now obtained
\begin{gather}
  {({\eta _t} + {\eta _x})_x} + \varepsilon {\left[ {\frac{3}{2}\eta {\eta _x}
  + \frac{1}{6}{\eta _{xxx}}} \right]_x}
  + \varepsilon {\delta ^2}\left[ {\frac{1}{2}{\eta _{yy}}} \right] +  \nonumber \\
  \mbox{} + {\varepsilon ^2}\frac{\partial}{\partial x}{\left[ { - \frac{3}{8}{\eta ^2}{\eta _x}
   + \frac{{23}}{{24}}{\eta _x}{\eta _{xx}} + \frac{5}{{12}}\eta {\eta _{xxx}}
   + \frac{{19}}{{360}}{\eta _{xxxxx}}} \right]} \nonumber \\
  \mbox{} + {\varepsilon ^2}{\delta ^2}\left[ \frac{9}{8}\eta _y^2 + \frac{1}{4}\eta {\eta _{yy}}
   + \frac{1}{4}{\eta _{xxyy}} - \frac{1}{2}{\eta _{xx}}\partial _x^{ - 2}({\eta _{yy}})\right.\nonumber\\
   \left. \mbox{} - \frac{3}{8}{\eta _x}\partial _x^{ - 1}({\eta _{yy}})+ {\eta _{xy}}\partial _x^{ - 1}({\eta _y})
   - \frac{{{\delta ^2}}}{8}\partial _x^{ - 2}({\eta _{yyyy}}) \right] = 0.
  \label{hkp}
\end{gather}
It can be seen that, in contrast to the $(1+1)$ dimensional case of the eKdV equation~(\ref{ekdv}),
the effect of dimensionality is now more pronounced. Indeed, the additional terms appearing,
measured by the parameter $\delta$, are highly nonlocal with the operator $\partial_x^{-1}$ being
applied multiple times. {As for the eKdV equation (\ref{ekdv}) this eKP equation has been derived in the specific case of water waves with the higher order $\epsilon^{2}$ coefficients taking specific values for this application.  For other applications, such as, again, internal waves in a stratified fluid, the higher order coefficients take other values specific to the physical application.  As for the ecKdV equation, a restricted form of the eKP equation with only higher order fifth order dispersion was derived by \cite{huang}.  It can be seen that the full eKP equation is much more extensive than just fifth order dispersion.}

\subsection{Surface tension}

As noted in the Introduction, the addition of capillary effects to the KdV equation can introduce
fundamental physical and mathematical changes in that the resulting incorporation of fifth order
dispersion can lead to resonance between solitary waves and undular bores and dispersive radiation
if the linear dispersion relation is non-convex \cite{patkdv,reskdv}. To include capillary effects,
only Bernoulli’s equation, the dynamic boundary condition, Eq.~(\ref{e:dynamic}), needs to be
altered through a curvature dependent surface tension term \cite{whitham_book,ablowitz_book}
\begin{eqnarray}
 && {\phi _t} + \frac{\varepsilon }{2}\left( {\phi _x^2 + {\delta ^2}\phi _y^2 + \frac{1}{{{\mu ^2}}}\phi _z^2} \right) + \eta
\nonumber \\
  &&= T\frac{{{\varepsilon ^2}{\eta _{xx}}(1 + {\varepsilon ^2}{\mu ^2}{\delta ^2}\eta _y^2)
  + {\mu ^2}{\delta ^2}{\eta _{yy}}(1 + {\varepsilon ^2}{\mu ^2}\eta _x^2)
  - 2{\varepsilon ^2}{\mu ^2}{\delta ^2}{\eta _{xy}}{\eta _x}{\eta _y}}}{{{{(1 + {\varepsilon ^2}\eta _x^2
  + {\varepsilon ^2}{\mu ^2}{\delta ^2}\eta _y^2)}^{3/2}}}}
  \nonumber \\
   &&= T\frac{{{\varepsilon ^2}{\eta _{xx}}(1 + {\varepsilon ^4}{\delta ^2}\eta _y^2)
   + {\varepsilon ^2}{\delta ^2}{\eta _{yy}}(1 + {\varepsilon ^3}\eta _x^2)
   - 2{\varepsilon ^4}{\delta ^2}{\eta _{xy}}{\eta _x}{\eta _y}}}{{{{(1 + {\varepsilon ^2}\eta _x^2
   + {\varepsilon ^4}{\delta ^2}\eta _y^2)}^{3/2}}}}
\end{eqnarray}
As such, it can be assumed that surface tension is a linear effect in the derivation of the KP
equation, as keeping terms that will affect the Euler equations at our level of approximation
requires the extra term $T(\eta_{xx} + \eta_{yy})\varepsilon^2+O(\varepsilon^4)$ in dimensionless
form \cite{ist,ablowitz_book,lamb}. The easiest manner in which to incorporate these capillary
effects in the KP equation is via the linear dispersion relation. The linear dispersion relation
for gravity-capillary waves is \cite{whitham_book}
\begin{align}
  {k_x}\omega  = {k_x}\sqrt {(k + T{k^3})\tanh (k)}  &= \left( {k_x^2
  + \frac{{ - 1 + 3T}}{6}k_x^4
  + \frac{{19 - 30T - 45{T^2}}}{{360}}k_x^6} \right)
  \nonumber \\
   &\mbox{} + {\delta ^2}\left( {\frac{1}{2} - \frac{{1 - 3T}}{4}k_x^2
   - \frac{{ - 19 + 30T + 45{T^2}}}{{144}}k_x^4} \right)k_y^2
   \nonumber \\
   &\mbox{} + {\delta ^4}\left( { - \frac{1}{{8k_x^2}} - \frac{{1 - 3T}}{{16}}
   - \frac{{ - 19 + 30T + 45{T^2}}}{{192}}k_x^2} \right)k_y^4.
  \label{disp2}
\end{align}
Then, using the usual connections $\omega \mapsto i\partial_t$, $k_{x} \mapsto -i\partial_{x}$ and
$k_{y} \mapsto -i\partial_{y}$ for linear dispersion relations \cite{whitham_book}, one may convert
the dispersion relation (\ref{disp2}) to the linearized version of the extended KP equation, which
is (\ref{hkp}) in the absence of surface tension under the KP weak transverse dispersion assumption
$|k_y/k_x| \ll 1$. Hence, we can deduce that the fully nonlinear version of the extended KP
equation, incorporating surface tension, is
\begin{gather}
  {({\eta _t} + {\eta _x})_x} + \varepsilon {\left[ {\frac{3}{2}\eta {\eta _x}
  + \frac{1}{6}(1-3T){\eta _{xxx}}} \right]_x}
  + \varepsilon {\delta ^2}\left[ {\frac{1}{2}{\eta _{yy}}} \right] +  \nonumber \\
\mbox{} + {\varepsilon ^2}{\left[ { - \frac{3}{8}{\eta ^2}{\eta _x}
   + \frac{{23}}{{24}}{\eta _x}{\eta _{xx}} + \frac{5}{{12}}\eta {\eta _{xxx}}
   + \frac{{19- 30T - 45{T^2}}}{{360}}{\eta _{xxxxx}}} \right]_x}
   \nonumber \\
\mbox{} + {\varepsilon ^2}{\delta ^2}\left[ \frac{9}{8}\eta _y^2 + \frac{1}{4}\eta {\eta _{yy}}
   + \frac{1}{4}(1-3T){\eta _{xxyy}} - \frac{1}{2}{\eta _{xx}}\partial _x^{ - 2}({\eta _{yy}})
   - \frac{3}{8}{\eta _x}\partial _x^{ - 1}({\eta _{yy}})\right.\nonumber\\
   \left.+ {\eta _{xy}}\partial _x^{ - 1}({\eta _y})
   - \frac{{{\delta ^2}}}{8}\partial _x^{ - 2}({\eta _{yyyy}}) \right] = 0.
  \label{hkpst}
\end{gather}
Note that other linear terms stemming from the dispersion relation (\ref{disp2}) and involving
surface tension do not appear in the eKP equation (\ref{hkpst}), but they would appear at a higher order of
approximation. In a similar manner, one may incorporate surface tension to the equations studied in Section \ref{s:1d}.

\subsection{The extended cylindrical Kadomtsev-Petviashvili equation}

The water wave equations~(\ref{e:laplace})--(\ref{e:kinematic}) can also be set
in cylindrical polar coordinates, namely
\begin{subequations}
\begin{gather}
  {\phi _{zz}} + \varepsilon \left( {{\phi _{rr}} + \frac{1}{r}{\phi _r} + \frac{{{\delta
  ^2}}}{{{r^2}}}{\phi _{\theta
  \theta }}} \right) = 0,
  \label{cLaplace} \\
  {\phi _z} = 0,\quad z =  - 1,
  \label{cbottom}\\
  {\phi _t} + \frac{\varepsilon }{2}\left( {\phi _r^2 + \frac{{{\delta ^2}}}{{{r^2}}}\phi _\theta ^2
  + \frac{1}{\varepsilon
  }\phi _z^2} \right) +
  \eta  = 0,\quad z = \varepsilon \eta,
  \label{cbernoulli} \\
  \varepsilon\left[{\eta _t} + \varepsilon \left( {{\phi _r}{\eta _r} + \frac{{{\delta ^2}}}{{{r^2}}}{\phi _\theta
  }{\eta _\theta }} \right)\right]
  = \phi _z,\quad z = \varepsilon \eta,
  \label{ckinematic}
\end{gather}
\end{subequations}
in order to derive the cylindrical KP (cKP) equation, also known as Johnson's equation
\cite{johnson2},
\[
{\frac{\partial}{\partial R}\left( {{H_T} + \frac{3}{2}H{H_R} + \frac{1}{6}{H_{RRR}} +
\frac{1}{{2T}}H} \right)}
  + \frac{{{\delta ^2}}}{{2{T^2}}}{H_{\Theta \Theta }}=0.
\]
Using the same approach, the extended cKP equation can be derived from the polar coordinate water wave
equations (\ref{cLaplace})--(\ref{ckinematic}).

We start by defining the new scaled and travelling wave variables
\[
R = \varepsilon (r - t), \quad T = {\varepsilon ^4}t, \quad \Theta  = \theta/ \varepsilon ^{3/2},
\quad \phi  = \varepsilon \Phi, \quad \eta  = {\varepsilon ^2}H,
\]
and expand the scaled velocity potential $\Phi$ as
\[
\Phi  = {\Phi _0} + {\varepsilon ^3}{\Phi _1} + {\varepsilon ^6}{\Phi _2}
+ {\varepsilon ^9}{\Phi _3} +\cdots
\]
This perturbation expansion can be used to solve Laplace's equation (\ref{cLaplace}). Substituting this
expansion into Laplace's equation and satisfying the bottom boundary condition (\ref{cbottom})
gives
\begin{gather*}
  {\Phi _0} = A(R,\Theta ,T),\quad {\Phi _1} =  - \frac{{{{\left( {z + 1} \right)}^2}}}{2}{A_{RR}},
   \\
  {\Phi _2} =  - \frac{{{{\left( {z + 1} \right)}^2}}}{{2T}}{A_R}
  + \frac{{{{\left( {z + 1} \right)}^4}}}{{24}}{A_{RRRR}}
  - {\delta^2}\frac{{{{\left( {z + 1} \right)}^2}}}{{2{T^2}}}{A_{\Theta \Theta }},
  \\
  {\Phi _3} = \frac{{{{\left( {z + 1} \right)}^2}R}}{{2{T^2}}}{A_R}
  + \frac{{{{\left( {z + 1} \right)}^4}}}{{12T}}{A_{RRR}}
  - \frac{{{{\left( {z + 1} \right)}^6}}}{{720}}{A_{RRRRRR}}\\
\mbox{}  + {\delta ^2}\frac{{{{\left( {z + 1} \right)}^2}R}}{{{T^3}}}{A_{\Theta \Theta }}
  + {\delta ^2}\frac{{{{\left({z + 1} \right)}^4}}}{{12{T^2}}}{A_{RR\Theta \Theta }}.
\end{gather*}
Next, we define $w=A_R$ and substitute into the dynamic boundary condition (\ref{cbernoulli})
(after differentiating with respect to $R$) and the kinematic boundary condition
(\ref{ckinematic}). This yields, to $O(\varepsilon^{6})$, the following equations
\begin{gather}
  {H_R} - {w_R} + {\varepsilon ^3}\left[ {\frac{2}{T}H - \frac{2}{T}w + \frac{{2R}}{T}{H_R}
  - \frac{{2R}}{T}{w_R} + {w_T} + w{w_R} + \frac{1}{2}{w_{RRR}}} \right] \nonumber\\
 \mbox{} + {\varepsilon ^6}\bigg[\frac{1}{T}{w^2} + \frac{{2R}}{T}w{w_R} + \frac{3}{{2T}}{w_{RR}} +
   \frac{R}{T}{w_{RRR}}
   + \frac{{2R}}{T}{w_T} + \frac{2}{T}\partial _R^{ - 1}({w_T}) \nonumber\\
 \mbox{} + \frac{{2R}}{{{T^2}}}H - \frac{{2R}}{{{T^2}}}w + \frac{{{R^2}}}{{{T^2}}}{H_R} -
   \frac{{{R^2}}}{{{T^2}}}{w_R}
   + \frac{{{\delta ^2}}}{{{T^2}}}{w_\Theta }\partial _R^{ - 1}({w_\Theta })
   + \frac{{{\delta ^2}}}{{2{T^2}}}{w_{R\Theta \Theta }} \nonumber \\
 \mbox{} + {H_R}{w_{RR}} + \frac{1}{2}{w_R}{w_{RR}} - \frac{1}{2}{w_{RRT}} + H{w_{RRR}}
   - \frac{1}{2}w{w_{RRR}} - \frac{1}{{24}}{w_{RRRRR}} \bigg] = 0,
   \label{cbernoulli2}
\end{gather}
and
\begin{gather}
   - {H_R} + {w_R} + {\varepsilon ^3}\left[ {\frac{1}{T}w - \frac{{2R}}{T}{H_R}
   + \frac{{2R}}{T}{w_R}
   + {H_T} + {{(Hw)}_R} - \frac{1}{6}{w_{RRR}}
   + \frac{{{\delta ^2}}}{{{T^2}}}\partial _R^{ - 1}({w_{\Theta \Theta }})} \right]\nonumber \\
   \mbox{} + {\varepsilon ^6}\bigg[\frac{1}{T}Hw + \frac{{2R}}{T}{H_T} + \frac{{2R}}{T}{H_R}w
   + \frac{{2R}}{T}H{w_R}
   - \frac{1}{{3T}}{w_{RR}} - \frac{R}{{3T}}{w_{RRR}} \nonumber \\
  \mbox{} + \frac{R}{{{T^2}}}w - \frac{{{R^2}}}{{{T^2}}}{H_R} + \frac{{{R^2}}}{{{T^2}}}{w_R}
   + \frac{{{\delta ^2}}}{{{T^2}}}{(H\partial _R^{ - 1}({w_\Theta }))_\Theta }
   - \frac{{{\delta ^2}}}{{3{T^2}}}{w_{R\Theta \Theta }} \nonumber \\
   \mbox{}- \frac{1}{2}{H_R}{w_{RR}} - \frac{1}{2}H{w_{RRR}} + \frac{1}{{120}}{w_{RRRRR}}\bigg] = 0.
   \label{ckinematic2}
\end{gather}
%
To make the above equations, (\ref{cbernoulli2}) and (\ref{ckinematic2}), compatible
we expand $w$ as
$$w = H + {\varepsilon ^3}({w_1} + {\delta ^2}{w_{12}}) + {\varepsilon ^6}({w_2} + {\delta
^2}{w_{22}}) + \ldots,$$
and then find that
\begin{align*}
  {w_1} &=  - \frac{1}{4}{H^2} + \frac{1}{3}{H_{RR}} - \frac{1}{{2T}}\partial _R^{ - 1}(H),\quad
  {w_{12}} =  - \frac{1}{{2{T^2}}}\partial _R^{ - 2}({H_{\Theta \Theta }}) \\
  {w_2} &= \frac{1}{8}{H^3} + \frac{3}{{16}}H_R^2 + \frac{1}{2}H{H_{RR}}
  + \frac{1}{{10}}{H_{RRRR}}
   + \frac{1}{T}\left[ {\frac{1}{6}{H_R} - \frac{1}{{16}}\partial _R^{ - 1}({H^2})} \right]\\
   &+ \frac{1}{{{T^2}}}\left[ {\frac{1}{2}\partial _R^{ - 1}(RH)
   + \frac{5}{8}\partial _R^{ - 2}(H)} \right], \\
  {w_{22}} &= \frac{1}{{2{T^2}}}\left[ {\frac{1}{3}{H_{\Theta \Theta }}
  + \frac{1}{4}\partial _R^{ - 2}(5H_\Theta ^2 + 2H{H_{\Theta \Theta }}
  + 3{H_R}\partial _R^{ - 1}({H_{\Theta \Theta }}))} \right]  \\
   & \mbox{} + \frac{1}{{{T^3}}}\left[ {\partial _R^{ - 2}(R{H_{\Theta \Theta }})
   + \frac{9}{4}\partial _R^{ - 3}({H_{\Theta \Theta \Theta }})} \right]
   + \frac{{3{\delta ^2}}}{{8{T^4}}}\partial _R^{ - 4}({H_{\Theta \Theta \Theta \Theta }}).
\end{align*}
Finally, substituting this expansion for $w$ into either equation~(\ref{cbernoulli2}) or (\ref{ckinematic2}) gives the extended cKP (ecKP) equation
\begin{gather}
  {\left[ {{H_T} + \frac{3}{2}H{H_R} + \frac{1}{6}{H_{RRR}} + \frac{1}{{2T}}H} \right]_R}
  + \frac{{{\delta ^2}}}{{2{T^2}}}{H_{\Theta \Theta }} \nonumber \\
  \mbox{} + {\varepsilon ^3}\frac{\partial}{\partial R}{\left[ { - \frac{3}{8}{H^2}{H_R} +
   \frac{{23}}{{24}}{H_R}{H_{RR}}
   + \frac{5}{{12}}H{H_{RRR}} + \frac{{19}}{{360}}{H_{RRRRR}}} \right]} \nonumber \\
   \mbox{} + \frac{{{\varepsilon ^3}}}{T}{\left[ {\frac{3}{{16}}{H^2} + \frac{1}{4}{H_{RR}}
   - \frac{1}{2}{H_R}\partial _R^{ - 1}(H)} \right]_R}
   + \frac{{{\varepsilon ^3}}}{{{T^2}}}\frac{\partial}{\partial R}{\left[ { - \frac{R}{2}H
   + \frac{1}{8}\partial _R^{ - 1}(H)} \right]} \nonumber \\
\mbox{} + {\varepsilon ^3}\frac{{{\delta ^2}}}{{{T^2}}}\left[ {\frac{9}{8}H_\theta ^2
   + \frac{1}{4}H{H_{\Theta \Theta }} - \frac{3}{8}{H_R}\partial _R^{ - 1}({H_{\Theta \Theta }})
   + {H_{R\Theta }}\partial _R^{ - 1}({H_\Theta }) - \frac{1}{2}{H_{RR}}\partial _R^{ -
   2}({H_{\Theta \Theta }})
   + \frac{1}{4}{H_{RR\Theta \Theta }}} \right] \nonumber \\
\mbox{} - {\varepsilon ^3}\frac{{{\delta ^2}}}{{{T^3}}}\left[ {  \frac{3}{4}\partial _R^{ -
   1}({H_{\Theta \Theta }})
   + {R}{{H}_{\Theta \Theta }}} \right]
   - {\varepsilon ^3}\frac{{{\delta ^4}}}{{{T^4}}}
   \partial _R^{ - 2}({H_{\Theta \Theta \Theta \Theta }}) = 0.
\end{gather}
Similarly to the cases of the extended cKdV and KP equations, the extended cKP equation is
also nonlocal due to the presence of the integral terms arising from the operators
$\partial^{-1}_{R}$ and $\partial^{-2}_{R}$. In addition,
as for the extended cKdV equation, as $T$ increases and the wave propagates outwards,
this equation approaches the eKdV equation. This is again expected since the expanding
wavefront becomes locally flat for large $T$.

\section{The Green-Naghdi equations}

Alternative approximations to the water wave equations to KdV-type weakly nonlinear, long wave
approximations are the Green-Naghdi (GN) and related equations. The GN equations provide a
depth averaged description of shallow water motion with a free surface under gravity. They are an
extension of the shallow water equations, which fully include the effects of finite fluid depth in
$\mu$, but are weakly dispersive. In the simplest case, the model was first derived by Serre
\cite{serre}, with extensions to two dimensions and the inclusion of higher order dispersion
subsequently derived \cite{matsuno3,khor}. The system of GN equations, as a
bidirectional, nonlinear dispersive wave model, has been proved to be a close approximation
to the 2D full water wave problem in that if the initial conditions for the
water wave equations and the GN equations are close, then the solutions of the two sets of
equations will remain close \cite{li}. In addition, it has been proved that if the dispersion parameter $\mu$
is small enough, solutions of the water wave equations and GN equations exist
for the same finite time.

We now focus on the higher order corrections to the Green-Naghdi equations. The higher
order correction to the GN system, while preserving the full nonlinearity of the original system,
along with its solitary wave solutions, was first obtained in Ref.~\cite{matsuno2}.  The 2D
Green-Naghdi shallow water model for surface gravity waves was extended to incorporate arbitrary higher order dispersive effects by \cite{matsuno3,matsuno}.
In addition, it was shown that the extended GN system has a Hamiltonian formulation \cite{matsuno2,matsuno3}, mirroring the
full water wave equations \cite{zakhham}.
Due to full nonlinearity, GN-type equations are valid above the KdV region of Figure~\ref{fig},
that of higher nonlinearity $\varepsilon$ and small dispersion $\mu$.

Let us first consider $(1+1)$ dimensional waves. We
introduce the mean horizontal velocity $\bar u=\bar u(x,t)$ by
\begin{equation}
\bar u(x,t)=\frac{1}{{\bar h}}\int_{-1}^{\varepsilon\eta}\phi_x(x,z,t)dz,
\quad {\bar h} = 1+\varepsilon\eta,
\label{ubar}
\end{equation}
where ${\bar h}$ is the total depth of the fluid, including the wave displacement. The horizontal
and vertical components of the surface velocity $u$ and $v$, respectively, are given in terms of
the velocity potential as
\begin{equation}
 u(x,t) = \phi_x(x,z,t) \quad \mbox{at} \quad z=\varepsilon\eta,
 \qquad
 v(x,t) = \phi_z(x,z,t) \quad \mbox{at} \quad z=\varepsilon\eta.
\label{velo}
\end{equation}
Multiplying the mean velocity expression (\ref{ubar}) by ${\bar h}$, differentiating the result
with respect to $x$, and then using Laplace's equation (\ref{e:laplace}) in the fluid bulk, the
bottom boundary condition (\ref{e:bottom}) and the velocity components (\ref{velo}), we obtain
\[
({\bar h}\bar u)_x=\varepsilon \eta_xu-v/\mu^2.
\]
Furthermore, as $\varepsilon\eta_x=\partial {\bar h}/\partial x$, we have
\begin{equation}
v=\mu^2\left[-\frac{\partial}{\partial x}({\bar h}\bar u) +\frac{\partial {\bar h}}{\partial x}u\right].
\label{v}
\end{equation}
Substitution of the expression~(\ref{v}) for the vertical velocity into the
kinematic boundary condition~(\ref{e:kinematic}) yields the evolution equation for the
total depth ${\bar h}$
\begin{equation}
\frac{\partial {\bar h}}{\partial t} +\varepsilon \frac{\partial}{\partial x}({\bar h}\bar u) =0.
\label{ht}
\end{equation}
An advantage of choosing the total depth ${\bar h}$ and the averaged horizontal velocity $\bar u$
as the dependent variables is that the mass conservation equation (\ref{ht}) is exact, meaning that
it does not involve any approximation.

Next, we differentiate the velocity components (\ref{velo}) with respect to $x$ and $t$ to obtain
\begin{subequations}
\begin{gather}
  u_x=\phi_{xx}+\varepsilon\phi_{xz}\eta_x,\quad u_t=\phi_{xt}+\varepsilon\phi_{xz}\eta_t,
  \label{ux}\\
  v_x=\phi_{xy}+\varepsilon\phi_{zz}\eta_x,\quad v_t=\phi_{zt}+\varepsilon\phi_{zz}\eta_t,
  \label{vx}
\end{gather}
\label{uxvx}
\end{subequations}
with all the derivatives evaluated at the surface $z=\varepsilon\eta$. Similarly,
\begin{equation}
\frac{\partial}{\partial x} \left(\frac{\partial}{\partial
t}\phi(x,\varepsilon\eta)\right)=\phi_{xt}+\varepsilon\phi_{zt}\eta_x.
\label{phit}
\end{equation}
Then, using Eqs.~(\ref{uxvx}) to eliminate $\phi_{xt}$ and $\phi_{zt}$, it is found that
Eq.~(\ref{phit}) reads
\begin{equation}
\frac{\partial}{\partial x} \left( \frac{\partial}{\partial t} \phi(x,\varepsilon \eta) \right)
=u_t+v_t {\bar h}_x-v_x {\bar h}_t.
\label{phit2}
\end{equation}
The final evolution equation for $u$ can now be obtained by differentiating the dynamic boundary
condition (\ref{e:dynamic}) with respect to $x$ and then using Eqs.~(\ref{v}), (\ref{ht}) and
(\ref{phit2}), to give
\begin{equation}
\frac{\partial u}{\partial t} + \frac{\partial v}{\partial t} \frac{\partial {\bar h}}{\partial x}
+ \varepsilon u \frac{\partial u}{\partial x}
+ \varepsilon u\frac{\partial {\bar h}}{\partial x} \frac{\partial
v}{\partial x} + \frac{\partial \eta}{\partial x} = 0.
\label{ut}
\end{equation}
Using the mass equation Eq.~(\ref{ht}) we may re-express this velocity equation as
\begin{align}
\frac{\partial}{\partial t} {\bar h}\left( u + v \frac{\partial {\bar h}}{\partial x} \right)
&+ \varepsilon \frac{\partial}{\partial x} {\bar h}
\left( u + v \frac{\partial {\bar h}}{\partial x} \right)
\nonumber \\&
\mbox{} + \varepsilon \left[
{\bar h}v \left( 2\frac{\partial {\bar h}}{\partial x} \frac{\partial \bar{u}}{\partial x}
+ {\bar h} \frac{\partial^{2}
\bar{u}}{\partial x^{2}} \right) + {\bar h} (u - \bar{u}) \left( \frac{\partial u}{\partial x} +
\frac{\partial v}{\partial x} \frac{\partial {\bar h}}{\partial x} \right) \right]
+{\bar h}\frac{\partial \eta}{\partial x} =0.
\label{ut2}
\end{align}
The system of equations (\ref{ht}) and (\ref{ut})--(\ref{ut2}) is equivalent to the basic water
wave equations (\ref{e:laplace})--(\ref{e:kinematic}). To obtain the extended GN equations, one
needs to express the velocities $u$ and $v$ in terms of the total depth ${\bar h}$ and the mean
horizontal velocity $\bar u$. To enable this change of variable, we set
\[
\bar u_t=\sum_{n=0}^\infty \mu^{2n}K_n,
\]
where $K_n$ are polynomials in ${\bar h}$. If one retains terms up to $O(\mu^{2n})$, the
extended GN equation, accurate up to $O(\mu^{2n})$, is obtained.

We now proceed to derive the extended GN equation explicitly for the case $n=2$ by truncating
the system of equations (\ref{ht}) and (\ref{ut2}) at $O(\mu^4)$.  As for the preceding extended
equations, we first solve Laplace's equation for the velocity potential $\phi$ subject to the
bottom boundary condition, and then find the mean velocity (\ref{ubar}) and velocity components
(\ref{velo}) accordingly as
\begin{align*}
\phi&=\sum_{n=0}^\infty(-1)^n\mu^{2n}\frac{(z+1)^{2n}}{(2n)!}\frac{\partial^{2n}A}{\partial
x^{2n}},\quad
  \bar u=\sum_{n=0}^\infty(-1)^n\mu^{2n}\frac{{\bar h}^{2n}}{(2n+1)!}
  \frac{\partial^{2n+1}A}{\partial x^{2n+1}},
  \\
u&= \sum_{n=0}^\infty(-1)^n\mu^{2n}\frac{{\bar h}^{2n}}{(2n)!}
\frac{\partial^{2n+1}A}{\partial x^{2n+1}},
  \quad
~~~v=\sum_{n=1}^\infty(-1)^n\mu^{2n}
\frac{{\bar h}^{2n-1}}{(2n-1)!}\frac{\partial^{2n}A}{\partial x^{2n}}.
\end{align*}
where, as above, $A=A(x,t)$. Retaining terms up to $O(\mu^4)$ for the mean horizontal velocity, we
have
\[
\bar u=A_x-\frac{\mu^2}{6}{\bar h}^2A_{xxx}+\frac{\mu^4}{120}{\bar h}^4A_{xxxxx}+O(\mu^6).
\]
This series may be inverted to obtain
\[
A_x=\bar u+\frac{\mu^2}{6}{\bar h}^2\bar u_{xx}+\mu^4\left[\frac{h^2}{36}\frac{\partial^{2}}{\partial
x^{2}}\left({\bar h}^2\bar u_{xx}\right)-\frac{{\bar h}^4}{120}\bar u_{xxxx}\right]
+O(\mu^6).
\]
Similarly,
\[
u=A_x-\frac{\mu^2}{2}{\bar h}^2A_{xxx}+\frac{\mu^4}{24}{\bar h}^4A_{xxxxx}+O(\mu^6),
\]
and after eliminating $A$ we obtain
\[
u=\bar u-\frac{\mu^2}{3}{\bar h}^2\bar u_{xx}-\mu^4\left[\frac{1}{45}{\bar h}^4\bar
u_{xxxx}+\frac{2}{9}{\bar h}^3\frac{\partial {\bar h}}{\partial x}
\bar u_{xxx}+\frac{1}{18}{\bar h}^2\frac{\partial^{2}{\bar h}^2}{\partial
x^{2}}\frac{\partial^{2}\bar{u}}{\partial x^{2}}\right]+O(\mu^6).
\]
Using this expression for the horizontal velocity, the expression for $v$, Eq.~(\ref{v}), can be
written as
\[
v=-\mu^2{\bar h}\bar u_x-\frac{\mu^4}{3}{\bar h}^2 {\bar h}_{x}\bar u_{xx}+O(\mu^6).
\]
The evolution equation for $\bar u$ follows by substituting these expressions for $u$ and $v$ into
Eq.~(\ref{ut}) and then using the mass equation Eq.~(\ref{ht}) to replace $\partial {\bar
h}/\partial t$. After some manipulations, in this manner we arrive at the following compact
equation for $\bar u$
\begin{align}
\bar u_t+\varepsilon\bar u\bar u_x+\eta_x&=\frac{\mu^2}{3{\bar h}}\frac{\partial}{\partial x}{\bar h}^3(\bar
u_{xt}+\varepsilon\bar u\bar
u_{xx}-\varepsilon\bar u_x^2) \nonumber\\
&+\frac{\mu^4}{45{\bar h}}\frac{\partial}{\partial x}\left[\frac{\partial}{\partial x}{\bar h}^5(\bar
u_{xxt}+\varepsilon\bar u\bar u_{xxx}-5\varepsilon \bar u_x\bar u_{xx})
-3\varepsilon {\bar h}^5\bar u_{xx}^2\right] + O(\mu^6).
\label{ubart}
\end{align}
The final system of equations (\ref{ht}) and (\ref{ubart}) is the extended version of the GN
equations, accurate up to $O(\mu^4)$.

Finally, these results can be extended to the two dimensional case and the 2D GN system reads
\begin{gather*}
\frac{\partial {\bar h}}{\partial t}+\varepsilon\nabla\cdot({\bar h}\bar{\bm u})=0, \\
{\bar{\bm u}}_t+\varepsilon(\bar{\bm u}\cdot\nabla){\bar{\bm
u}}+\nabla\eta=\mu^2R_1+\mu^4R_2+O(\mu^6),
\end{gather*}
where
\begin{align*}
R_1&=\frac{1}{3{\bar h}}\nabla\left[{\bar h}^3\{\nabla\cdot{\bar{\bm u}}_t+\varepsilon(\bar{\bm
u}\cdot\nabla)(\nabla\cdot\bar{\bm u})
-\varepsilon(\nabla\cdot\bar{\bm u})^2\}\right],
\\
R_2&=\frac{1}{45{\bar h}}\nabla\Bigl[\nabla\cdot\bigl\{{\bar h}^5\nabla(\nabla\cdot{\bar{\bm u}}_t)
+\varepsilon {\bar h}^5(\nabla^2(\nabla\cdot\bar{\bm u}))\bar{\bm u}-5\varepsilon {\bar h}^5(\nabla\cdot\bar{\bm
u})\nabla(\nabla\cdot\bar{\bm u}) \\
& \mbox{} +\varepsilon\nabla {\bar h}^5\times(\bar{\bm u}\times\nabla(\nabla\cdot\bar{\bm u}))\bigr\}
-2\varepsilon {\bar h}^5\{\nabla(\nabla\cdot\bar{\bm u})\}^2\Bigr]\\
& \mbox{} -\frac{\varepsilon}{45{\bar h}}\Bigl[\nabla\cdot\{{\bar h}^5\nabla(\nabla\cdot\bar{\bm
u})\}\nabla(\nabla\cdot\bar{\bm u})
+\frac{1}{2}{\bar h}^5\nabla\{\nabla(\nabla\cdot\bar{\bm u})\}^2\Bigr].
\end{align*}
Note here we have taken $\delta=1$ in Eq.\ (\ref{e:laplace}).

\section{Conclusions}

The standard weakly nonlinear, long wave approximations to the water wave equations,
the Boussinesq system and the KdV equation in $(1+1)$ dimensions and the KP equation
in $(2+1)$ dimensions, are asymptotic approximations one order beyond linear waves
for which weak nonlinearity is in balance with weak dispersion. As outlined in the Introduction,
these equations appear generically as asymptotic reductions of other models
(such as the NLS equation, discrete dynamical lattices, etc) that play a key role
in a wide range of physical contexts. Moreover, in a variety of applications, it has been found
that it is necessary to extend these asymptotic equations to the next order, so that
higher amplitude and steeper waves can be modelled. In addition, it has been found
that such extensions are necessary in order to capture effects not included at the KdV order,
such as resonance. While the extended equations derived here, in particular the eKdV, ecKdV
and eKP equations, have been derived from the water wave equations, they have much a much wider
applicability, as noted in the Introduction.

In the present work, extended weakly nonlinear, weakly dispersive reductions of the water wave equations in both $(1+1)$ and $(2+1)$ dimensions have been derived, as
well as extended versions of the fully nonlinear, weakly dispersive GN equations. These extended
equations have been derived using asymptotic analyses which highlight the connections between the
various equations. From a mathematical point of view, the extended equations are quite relevant
to important notions, such as the Hamiltonian structure, as well as the integrability and asymptotic integrability, of weakly nonlinear dispersive wave equations.
It is hoped that the newly derived equations will  prove to be useful in theoretical contexts as, in existing work, the additional higher order terms were added on an ad hoc basis. Furthermore, it is anticipated that this derivation and review of higher order weakly
dispersive, weakly nonlinear equations will prove useful for modelling waves in fluids, plasmas,
nonlinear optics, and other application areas.
It is anticipated that the newly derived equations will prove to be useful in theoretical contexts as, in existing work, the additional higher order terms were added on an
{\it ad hoc} basis.

{From a mathematical point of view, important themes for future studies concern the properties and the solutions of the extended equations.  In that regard, an interesting question is if the extended equations have a Hamiltonian structure, as is the case for the extended GN equations \cite{matsuno2,matsuno3}.  For instance, the eKdV equation seems to lack the Hamiltonian property
as an exact energy conservation equation for it has not been found. Furthermore, resonance effects \cite{ekdv} strongly suggest that a Hamiltonian formulation of the eKdV equation cannot be constructed. More generally, the existence of
conserved quantities for the extended equations, which is a ``stepping stone'' to complete
integrability, is an important problem.  A promising way to investigate these issues is through the notion of {\it asymptotic integrability} \cite{hkdvbore}.  In this context, there exists a map, much like the one connecting the functions $w$ and $\eta$ of Sections \ref{s:ekdv} and \ref{s:ekp}, which is constructed to eliminate the higher order contributions (by moving them to higher than the required order) \cite{hkdvbore}.  In this manner, the higher order equation can be reduced to its integrable counterpart at a given asymptotic order.  Then, the solution of the integrable equation is used to construct, up to the required order in $\varepsilon$, the solution of original higher-order equation.}

It is hoped that this work will inspire theoretical studies in this direction, which may also
concern the asymptotic integrability of other higher order weakly nonlinear dispersive wave
equations. Furthermore, it is anticipated that this derivation and review of higher order weakly
dispersive, weakly nonlinear equations will prove useful for modelling waves in fluids, plasmas,
nonlinear optics, and other application areas.


\end{document}